\documentclass[twocolumn,showpacs,preprintnumbers,amsmath,amssymb]{revtex4}
\usepackage{graphicx}
\usepackage{dcolumn}
\usepackage{bm}
\usepackage{dpscolor}
\usepackage{epsfig}

\begin{document}


\title{
Virtual Compton Scattering measurements in the $\gamma^*
N\rightarrow \Delta$ transition.}

\author{N.F.~Sparveris$^1$\footnote{current address: Massachusetts Institute of Technology},
P.~Achenbach$^2$, C.~Ayerbe Gayoso$^2$, D.~Baumann$^2$,
J.~Bernauer$^2$, A.M.~Bernstein$^4$, R.~B\"ohm$^2$, D.~Bosnar$^5$,
T.~Botto$^4$, A.~Christopoulou$^1$, D.~Dale$^6$\footnote{current
address: Idaho State University, Department of Physics, Pocatello,
Idaho 83209, USA}, M.~Ding$^2$, M.O.~Distler$^2$, L.~Doria$^2$,
J.~Friedrich$^2$, A.~Karabarbounis$^1$, M.~Makek$^5$, H.~Merkel$^2$,
U.~M\"uller$^2$, I.~Nakagawa$^3$, R.~Neuhausen$^2$,
L.~Nungesser$^2$, C.N.~Papanicolas$^1$\footnote{corresponding
author, e-mail address: cnp@iasa.gr}, B.~Pasquini$^8$,
A.~Piegsa$^2$, J.~Pochodzalla$^2$, M.~Potokar$^7$, M.~Seimetz$^2$,
S.~\v Sirca$^7$, S.~Stave$^4$\footnote{current address: Triangle
Universities Nuclear Laboratory, Duke University, Durham, North
Carolina 27708, USA}, S.~Stiliaris$^1$, Th.~Walcher$^2$ and
M.~Weis$^2$}

\affiliation{$^1$Institute of Accelerating Systems and Applications
and Department of Physics, University of Athens, Athens, Greece}

\affiliation{$^2$Institut fur Kernphysik, Universit\"at Mainz,
Mainz, Germany}

\affiliation{$^3$Radiation Laboratory, RIKEN, 2-1 Hirosawa, Wako,
Saitama 351-0198, Japan}

\affiliation{$^4$Department of Physics, Laboratory for Nuclear
Science and Bates Linear Accelerator Center,
\\ Massachusetts Institute of Technology, Cambridge, Massachusetts 02139}

\affiliation{$^5$Department of Physics, University of Zagreb,
Croatia}

\affiliation{$^6$Department of Physics and Astronomy, University of
Kentucky, Lexington, Kentucky 40206 USA}

\affiliation{$^7$Institute Jo\v zef Stefan, University of Ljubljana,
Ljubljana, Slovenia}

\affiliation{$^8$Dipartimento di Fisica Nucleare e Teorica,
Universit\`a degli Studi di Pavia, and INFN, Sezione di Pavia, Pavia,
Italy}

\date{\today}

\begin{abstract}
We report on new H$(e,e^\prime p)\gamma$ measurements in the
$\Delta(1232)$ resonance at $Q^2=0.06$ (GeV/c)$^2$ carried out
simultaneously with H$(e,e^\prime p)\pi^0$. It is the lowest $Q^2$
for which the virtual Compton scattering (VCS) reaction has been
studied in the first resonance region. The VCS measured cross
sections are well described by dispersion-relation calculations in
which the multipole amplitudes derived from  H$(e,e^\prime p)\pi^0$
data are used as input, thus confirming the compatibility of the
results. The derived resonant magnetic dipole amplitude
$M^{3/2}_{1+} = (40.60 \pm 0.70_{stat+sys})(10^{-3}/m_{\pi^+})$ at
$W=$ 1232 MeV is in excellent agreement with the value extracted
from H$(e,e^\prime p)\pi^0$  measurements.

\end{abstract}


\maketitle

During the past three decades an extensive effort has been devoted
to the study of the $\gamma^* N\rightarrow \Delta$ transition in
order to precisely determine the resonant amplitudes involved in
the process \cite{PaBe07}. According to spin-parity selection
rules, only magnetic dipole $(M^{3/2}_{1+})$ and electric
quadrupole $(E^{3/2}_{1+})$ or Coulomb quadrupole
$(S^{3/2}_{1+})$ multipoles are allowed to contribute to this
transition. The magnetic dipole amplitude $M^{3/2}_{1+}$
dominates, a manifestation of the spin flip character of the
transition.  The presence of quadrupole amplitudes identifies and
helps elucidate the origins of the non-spherical components in
the nucleon wavefunction
\cite{is82,pho2,pho1,frol,pos01,merve,bart,Buuren,joo,spaprc,kun00,spaprl,kelly,stave,ungaro,dina,sato,dmt00,kama,mai00,said,elsner,spaplb,rev,pvy}.
It is the complex quark-gluon and meson cloud dynamics of hadrons
that give rise to non-spherical components in their wavefunction
which at the classical limit and at large wavelengths will
correspond to a ``deformation''.
\begin{figure}
\centerline{\psfig{figure=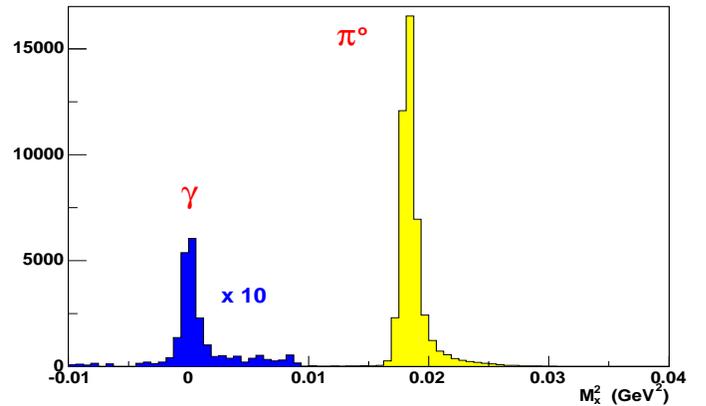,width=10.0cm,height=6.0cm}}
\smallskip
\caption{(Color online) The derived missing mass spectrum, plotted
as a function of the square of missing mass shows the superb
resolution achieved, essential to isolating the small photon decay
branch of the $\Delta(1232)$ Resonance. Channels for $M_x^2 < 0.01
GeV^2$ have been multiplied by a factor of ten.} \label{fig:mm1}
\end{figure}


\begin{figure*}[tbc]
\centerline{\psfig{figure=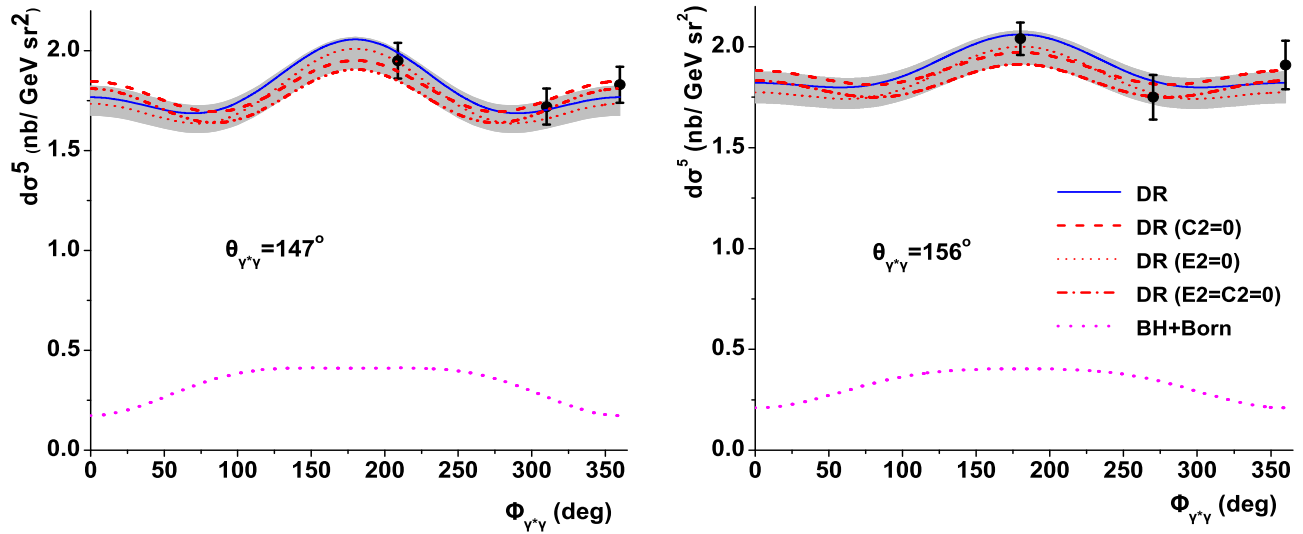,width=19.0cm,height=8.8 cm}}
\smallskip
\caption{(Color online) Comparison of the measured VCS cross
sections at $\theta_{\gamma^*\gamma}=147^\circ$  and $156^\circ$
 as a function of the azimuthal angle
$\phi_{\gamma^*\gamma}$ to Dispersion Relation (DR) calculations
\cite{dr} with different input.
 a)Solid curves: input from the standard solutions of MAID 2003;
b)dashed curves: As in a)  but with $S^{3/2}_{1+}=0$; c)dotted
curves: As in a) with $E^{3/2}_{1+}=0$; d)dashed-dotted curves: DR
results with $S^{3/2}_{1+}=E^{3/2}_{1+}=0$. The shaded band is
derived when the dipole and quadrupole amplitudes are constrained to
the values extracted from the simultaneously measured H$(e,e^\prime
p)\pi^0$  \cite{stave}. The DR calculation with the parameters
$\Lambda_{\alpha}=1.0$ GeV and $\Lambda_{\beta}=0.6$ GeV
(dash-dot-dot curves) is also shown with $\Lambda_{\alpha}=0.7$ GeV
and $\Lambda_{\beta}=0.5$ GeV (solid curves). } \label{fig:vcs22}
\end{figure*}

Up to now only the dominant  H$(e,e^\prime p)\pi^0$ and the
H$(e,e^\prime \pi^+)n$ $\Delta$ reaction channels, with branching
ratios of approximately 66$\%$ and 33$\%$ respectively, have been
utilized for the determination of the resonant amplitudes in the
transition. In this work, we present results obtained for the
first time from the weak H$(e,e^\prime p)\gamma$ channel. The
different nature of this reaction channel, being purely
electromagnetic, and the fact that it was measured simultaneously
with the dominant H$(e,e^\prime p)\pi^0$ channel  \cite{stave},
allows for important tests of the reaction framework and of the
systematic uncertainties of the extracted resonant amplitudes.
The measurement was made possible because of the excellent
quality of the MAMI beam and the superb resolution and wide
acceptance of its spectrometers which yielded a high resolution
missing mass spectrum with the $\gamma$ and $\pi^0$
simultaneously measured and widely separated (see Fig.1).

The magnetic dipole amplitude is accurately known and the
existence of non-spherical components in the nucleon wavefunction
has been established through the extraction of resonant quadrupole
amplitudes in the pion-electroproduction channels \cite{rev}.
However, the control and quantification of the model error of the
resonant amplitudes as well as the understanding and significance
of the various interfering channels in the transition ("background
channels") are still open issues and of major importance. The
exploitation of the H$(e,e^\prime p)\gamma$ reaction channel can
be of central importance towards this direction. The resonant
amplitude contributions are isolated within different theoretical
frameworks in the pion- and photon-electroproduction channels.
Background contributions, which need to be known for the
determination of the weak quadrupole resonant amplitudes, are of
different nature for these channels and therefore present
different theoretical problems. Thus an important cross check on
the model dependence of the analysis is offered through the
comparison of the results obtained from the photon- and the
pion-electroproduction reactions.

In pion-electroproduction  the reaction cross section can be
factorized into a virtual photon flux and a sum of partial cross
sections ($\sigma_{T},\sigma_{L},\sigma_{LT}$, $\sigma_{TT}$,
$\sigma_{LT'}$) which contain the physics of interest. This is not
possible in the case of photon-electroproduction because the
detected photon can emerge not only from the de-excitation of the
$\Delta(1232)$ but also from the incoming or scattered electron,
from the Bethe-Heitler (BH) process. The  VCS reaction
$\gamma^*p\rightarrow \gamma p$ amplitude also contains a Born
component. The BH and Born contributions are well understood and
precisely calculable with the nucleon electromagnetic form factors
as inputs. The non-Born amplitude contains the physics of interest,
which includes resonant amplitudes as well as Generalized
Polarizabilities (GPs).
 Previous experiments
\cite{vcs1,vcs2,vcs3} have  focused either on the extraction of
the GPs from cross section measurements or the study of the
imaginary part of the VCS amplitude  through  beam helicity
asymmetry measurements~\cite{vcs4}. In this work sensitivities to
the resonant amplitudes and to the GPs in the
$\gamma^*p\rightarrow \Delta $ have been explored for the first
time through cross section measurements.

\begin{figure}
\centerline{\psfig{figure=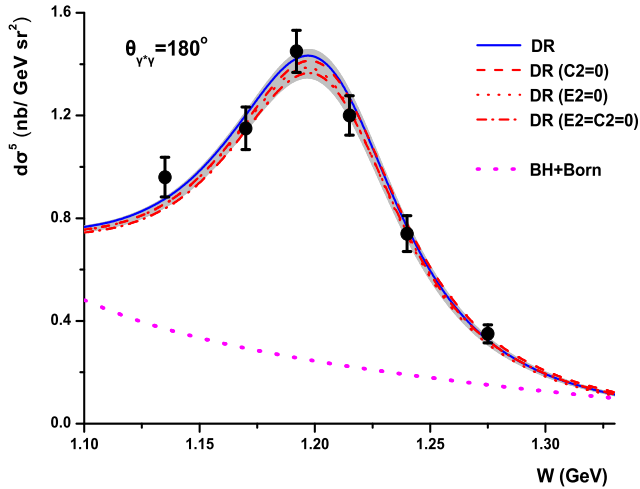,width=9.5cm,height=7.6 cm}}
\smallskip
\caption{(Color online) The measured VCS cross sections at
$\theta_{\gamma^*\gamma}=180^\circ$ as a function of W. The labeling
conventions are the same as in Fig.~2.} \label{fig:vcs3}
\end{figure}

\begin{figure}
\centerline{\psfig{figure=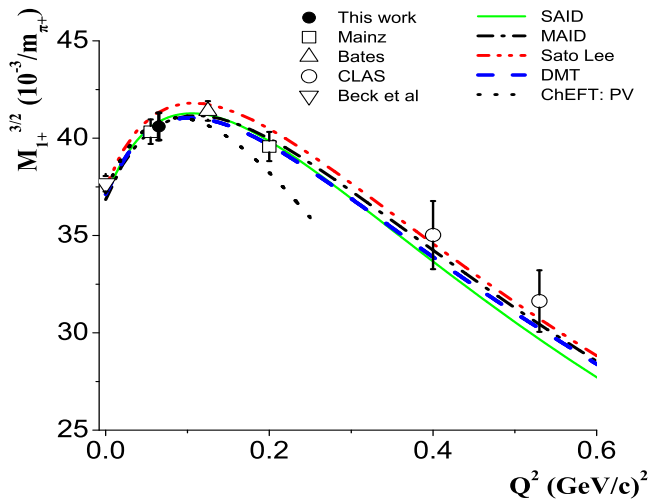,width=9.5cm,height=7.6 cm}}
\smallskip
\caption{ (Color online) Results for $M_{1+}^{3/2}$  as a function
of $Q^2$ from this work (VCS) and from previous
pion-electroproduction experiments
\cite{pho1,joo,spaprl,stave,spaplb} (errors include statistical and
systematic uncertainties). The two results at $Q^2=0.06$ (GeV/c)$^2$
have been shifted by 0.005 (GeV/c)$^2$ in order to be
distinguishable. The theoretical predictions of MAID \cite{mai00},
DMT \cite{dmt00,kama}, SAID \cite{said}, Sato-Lee \cite{sato} and
the ChEFT of Pascalutsa-Vanderhaeghen \cite{pv} are also shown.}
 \label{fig:vcs4}
\end{figure}

The experiment was performed at the Mainz Microtron using the A1
magnetic spectrometers \cite{spectr}. The experimental arrangement
and parameters are those  reported in \cite{stave}. An 855 MeV
electron beam with an average beam current of  25~$ \mu A$ was
employed on a liquid-hydrogen target. Electrons and protons were
detected in coincidence with spectrometers A and B respectively. The
H$(e,e^\prime p)\gamma$ reaction was performed at $Q^2=0.06$
(GeV/c)$^2$ and at the top of the $\Delta(1232)$ resonance. The
measurements were taken  for $\theta_{\gamma^*\gamma}$ values from
$147^\circ$ up to $180^\circ$, with $\theta_{\gamma^*\gamma}$ the
polar angle in the c.m. frame between the initial and final photons
of the VCS process, and for a range of azimuthal angles with respect
to the electron scattering plane $\phi_{\gamma^*\gamma}$ from
$180^\circ$ to $360^\circ$. Parallel cross section measurements at
$\theta_{\gamma^*\gamma}=180^\circ$ have also been performed
covering a range of the $p\gamma^*$ c.m. energy $W$ from $1135$ MeV
to $1275$ MeV.

To facilitate cross section extraction the data were sorted in
kinematic bins of the following widths: $ \Delta W = 15$ MeV,
$\Delta Q^2= 0.014$ (GeV/c)$^2$, $\Delta \theta_{\gamma^*\gamma
}=8^\circ$ and $\Delta \phi_{\gamma^*\gamma}= 30^\circ$. Point
cross sections were derived from the finite acceptances by
projecting the measured values using the Dispersion-Relation (DR)
 model~\cite{dr}. In this model the VCS non-Born
contribution is given in terms of dispersive integrals relating
the real and imaginary parts of the amplitude. The imaginary part
is calculated, through the unitarity relation, from the scattering
amplitudes of electro- and photo-production on the nucleon,
taking into account the dominant contribution from $\pi N$
intermediate states. The DR model has two free parameters,
$\Lambda_{\alpha}$ and $\Lambda_{\beta},$ related to the dipole
electric and magnetic GPs, respectively, while the amplitudes for
$\gamma^{(*)}p\rightarrow \gamma \pi$ entering the unitarity
relation are taken from MAID 2003~\cite{mai00}.
 The projection to the central kinematical values
introduces an uncertainty of the order of $1\%$ to the derived cross
section values. Radiative corrections were applied to the data using
Monte Carlo simulation; a detailed description of these corrections
can be found in \cite{radcorr}. Elastic scattering data from H and
$^{12}$C for calibration purposes were taken at 600 MeV. The
systematic uncertainties of the cross sections have been determined
to be about 4$\%$, whereas the statistical uncertainty is about
3$\%$ to 4$\%$ depending on the kinematics; thus both the systematic
and the statistical errors have an equivalent contribution to the
extracted cross section uncertainties.

In Fig.~2 we present the experimental results for the extracted
cross sections at $\theta_{\gamma^*\gamma} = 156^\circ$ and
$147^\circ$ for the measured azimuthal  angles
$\phi_{\gamma^*\gamma}$. In Fig.~3 the cross sections measured at
$\theta_{\gamma^*\gamma}= 180^\circ$ as a function of the
invariant mass {\it W} are presented. The depicted uncertainties
result from the combination of statistical and systematic errors
added in quadrature. The experimental results are compared with
the predictions of the DR model with input from MAID 2003 for the
magnetic dipole and the electric and coulomb quadrupole
amplitudes. The sensitivity to the quadrupole amplitudes has been
explored in the three ``spherical solutions" where either one or
both quadrupole amplitudes are set equal to zero. The sensitivity
of the GPs has been explored by varying the $\Lambda_{\alpha}$ and
$\Lambda_{\beta}$ parameters (mass scale free parameters that
determine the $Q^2$ dependence of the polarizabilities \cite{dr});
it was found to be insignificant compared to that of the resonant
amplitudes.

The data reported here allow for the first time  to compare
directly the results of both VCS and pion electro-production
channels. This is achieved by using as input in the DR
calculation the amplitudes (values and the uncertainties)
extracted from the pion electro-production channel \cite{stave}.
The shaded band in Fig.~2 and Fig.~3 shows the allowed range of
compatibility (1$\sigma$) which is driven by the uncertainty in
the magnetic dipole amplitude. By comparing the extracted VCS
cross sections and the shaded band constrained by the pion
electro-production measurements one has a direct cross check of
both reaction channels. The two reactions measure the same
physical signal within different physical background, and
therefore the compatibility of the derived amplitudes (or lack
there of) tests their consistency and reliability. As shown in
Fig.~2 and Fig.~3 the VCS results are in excellent agreement with
the solutions compatible with the pion-electroproduction
measurements.

The statistical accuracy of the cross section measurements allows
the accurate extraction of the resonant magnetic dipole amplitude
but not that of the two quadrupole multipoles. The $M^{3/2}_{1+}$
value extracted from the VCS data, within the framework of the
Dispersion Relations calculation, of $(40.60 \pm
0.70_{stat+sys})(10^{-3}/m_{\pi^+})$ is in good agreement with
the value of $(40.33 \pm 0.63_{stat+sys} \pm 0.61_{model}
)(10^{-3}/m_{\pi^+})$ determined through the pion
electro-production channel \cite{stave} and of comparable
precision. The derived values are in agreement with theoretical
predictions and the overall trend of the existing data as shown
in Fig.~4.

In summary, we have presented measurements of the VCS reaction in
the $\Delta(1232)$ resonance region obtained simultaneously with
the dominant pion electro-production channel at the low momentum
transfer of $Q^2=0.06$ (GeV/c)$^2$. Cross sections have been
extracted on the top and at the wings of the resonance, both in-
and out-of-plane. The measured cross sections are found to be
exceedingly well described by Dispersion Relation calculations.
The sensitivity of the data to the resonant dipole and quadrupole
amplitudes as well as to the GPs has been explored. At the
kinematics of the measurement the effect of the GPs is
inconsequential. Higher sensitivity is exhibited to the
quadrupole amplitudes but the large statistical uncertainty of the
cross sections does not allow for their accurate separation.
Given that these results were obtained in short acquisition times
optimized for the $\pi^0$ channel, future dedicated measurements
of higher statistical accuracy and with an improved control of the
systematic error could provide better sensitivity to the
quadrupole amplitudes. Covering  a wider range of proton angles
will also be of value in improving the sensitivity to the
resonant amplitudes. The VCS cross sections are in excellent
agreement with the resonant amplitudes solution from the
pion-electroproduction measurements~\cite{stave}, which were
obtained simultaneously. This first comparison between the
results from photo- and pion-electroproduction channels, provides
a stringent cross check for the extraction of the resonance
multipole amplitudes, rendering further support  to the
conclusions drawn previously validating the conjectured
deformation of the Nucleon ~\cite{spaprl,stave, rev}.

We would like to thank the MAMI accelerator group for providing
the excellent beam quality required for these demanding
measurements. This work is supported at Mainz by the
Sonderforschungsbereich 443 of the Deutsche
Forschungsgemeinschaft (DFG) and by the program PYTHAGORAS
co-funded by the European Social Fund and National Resources
(EPEAEK II).


\begin{references}

\bibitem{PaBe07} C. N. Papanicolas  and A. M. Bernstein, AIP Conference Proceedings
104, 1 (2007) and articles therein.

\bibitem{kun00} C.~Kunz {\it et al.}, Phys. Lett. B {\bf 564}, 21 (2003).

\bibitem{spaprl} N.F.~Sparveris {\it et al.}, Phys. Rev. Lett. {\bf
94}, 022003 (2005).

\bibitem{is82} N. Isgur, G. Karl and R. Koniuk, {\it Phys. Rev.} D {\bf 25}, 2394 (1982);
S. Capstick and G. Karl, {\it Phys. Rev.}  D {\bf 41}, 2767 (1990).

\bibitem{pho2} G.~Blanpied {\it et al.}, Phys. Rev. Lett. {\bf 79}, 4337 (1997).

\bibitem{pho1} R.~Beck {\it et al.}, Phys. Rev. C {\bf 61}, 35204 (2000).

\bibitem{frol} V.V.~Frolov {\it et al.}, Phys. Rev. Lett. {\bf 82}, 45 (1999).

\bibitem{pos01} T. Pospischil {\it et al.}, Phys. Rev. Lett. {\bf 86} (2001), 2959.

\bibitem{merve} C.~Mertz {\it et al.}, Phys. Rev. Lett. {\bf 86}, 2963 (2001).

\bibitem{bart} P.~Bartsch {\it et al.}, Phys. Rev. Lett. {\bf 88}, 142001 (2002).

\bibitem{Buuren} L.D.~van Buuren {\it et al.}, Phys. Rev. Lett. {\bf 89}, 12001 (2002).

\bibitem{joo} K.~Joo {\it et al.}  Phys. Rev. C {\bf 68}, 032201 (2003); Phys. Rev. C {\bf 70}, 042201
(2004).



\bibitem{spaprc} N.F.~Sparveris {\it et al.}, Phys. Rev. {\bf C67}, 058201 (2003).

\bibitem{kelly} J.J.~Kelly {\it et al.}, Phys. Rev.
Lett. {\bf 95}, 102001 (2005).

\bibitem{stave} S.~Stave {\it et al.}, Eur. Phys.
J. A {\bf  30}, 471 (2006).

\bibitem{ungaro}   M.~Ungaro {\it et al.}, Phys. Rev. Lett. {\bf
97}, 112003 (2006).

\bibitem{dina} C.~Alexandrou {\it et al.}, Phys. Rev D {\bf 69} 114506
(2004); Phys. Rev. Lett. {\bf 94}, 021601 (2005) and
arXiv:0710.4621 [hep-lat].

\bibitem{sato} T.~Sato and T.-S.H.~Lee, Phys. Rev. C {\bf 63}, 055201 (2001).

\bibitem{dmt00} S.S. Kamalov and S. Yang, Phys. Rev. Lett. {\bf 83}, 4494
(1999).

\bibitem{kama} S.S. Kamalov {\it et al.}, Phys. Lett. B {\bf  522}, 27 (2001).

\bibitem{mai00} D.~Drechsel {\it et al.}, Nucl. Phys. A {\bf  645}, 145 (1999).


\bibitem{said} R.A.~Arndt, {\it et al.} Phys. Rev. C {\bf 66}, 055213 (2002); nucl-th/0301068 and http://gwdac.phys.gwu.edu

\bibitem{elsner} D.~Elsner {\it et al.}, Eur. Phys. J. A {\bf  27} 91-97 (2006).

\bibitem{spaplb} N.F. Sparveris {\it et al.},Phys. Lett. {\bf B 651}, 102 (2007).


\bibitem{rev} A.M. Bernstein and C.N. Papanicolas, AIP Conf. Proc. 904: 1-22 (2007).
\bibitem{pvy} V.~Pascalutsa, M. Vanderhaeghen, S.-N.~ Yang, Phys. Rept.
{\bf 437}, 125 (2007).

\bibitem{vcs1} J.~Roche {\it et al.}, Phys. Rev. Lett. {\bf 85}, 708 (2000).

\bibitem{vcs2} G. Laveissiere {\it et al.}, Phys. Rev. Lett. {\bf
93}, 122001 (2004).

\bibitem{vcs3} P. Bourgeois {\it et al.}, Phys. Rev. Lett. {\bf 97}, 212001 (2006) ; \\
P. Bourgeois, Ph.D thesis, Univ. of Massachusetts, 2005.

\bibitem{vcs4} I.K. Bensafa {\it et al.}, Eur. Phys. J. A {\bf 32},69 (2007).

\bibitem{pv} V.~Pascalutsa and M.~Vanderhaeghen, Phys. Rev. Lett. {\bf 95}, 232001 (2005);
Phys. Rev. D {\bf 73}, 034003 (2006).

\bibitem{spectr} K.I.~Blomqvist {\it et al.}, Nucl. Instrum. Methods A {\bf  403}, 263 (1998).

\bibitem{dr} B.~Pasquini {\it et al.}, Eur. Phys. J. A {\bf 11}, 185 (2001);
D.~Drechsel, et al., Phys. Rep. {\bf 378}, 99 (2003).

\bibitem{radcorr} M.~Vanderhaeghen  {\it et al.}, Phys. Rev. {\bf C62}, 025501 (2000).








\end{references}
\end{document}